\documentclass[MSNbibl,number,dvips]{arxstspdf}
\usepackage{flushend}
\usepackage{stfloats}
\usepackage{url,breakurl}
\volume{29}
\issue{2}
\pubyear{2014}
\firstpage{165}
\lastpage{166}
\doi{10.1214/14-STS481} 



\begin{document}
\begin{frontmatter}

\title{Four Papers on Contemporary Software Design Strategies for
Statistical Methodologists}
\runtitle{Introduction to Symposium on Software Design}

\begin{aug}
\author[a]{\fnms{Vincent}~\snm{Carey}\corref{}\ead[label=e1]{stvjc@channing.harvard.edu}}
\and
\author[b]{\fnms{Dianne}~\snm{Cook}\ead[label=e2]{dicook@iastate.edu}}
\runauthor{V. Carey and D. Cook}

\affiliation{Harvard Medical School and Iowa State University}

\address[a]{Vincent Carey is Associate Professor of Medicine (Biostatistics), Channing Division of Network
Medicine, Harvard Medical School, Boston, Massachusetts 02115, USA \printead{e1}.}
\address[b]{Dianne Cook is Professor, Department of Statistics, Iowa State University, Ames, Iowa 50011,
USA \printead{e2}.}
\end{aug}


\end{frontmatter}

Software design impacts much of statistical analysis and, as technology
changes, dramatically so in recent years, it is exciting to
learn how statistical software is adapting and changing. This leads
to the collection of papers published here, written by John
Chambers, Duncan Temple Lang, Michael Lawrence, Martin Morgan,
Yihui Xie, Heike Hofmann and Xiaoyue Cheng.

John Chambers has been at the forefront of advances in
computing for data analysis since the 1970s, and his contributions
have been recognized through the 1998 Association for Computing
Machinery Software Systems Award. The award statement
noted that the S system
``forever altered how people analyze, visualize,
and manipulate data'' and remarked on the role of Chambers'
``insight, taste, and effort'' in establishing
S as an ``elegant, widely accepted, and enduring
software system.'' Chambers' contribution to this symposium
includes historical background and focuses on the
joint roles of object-oriented and functional programming
disciplines in fostering effective and extensible data
analysis environments.

Duncan Temple Lang is the creator of the Omegahat
project \cite{omega} and a long-standing member of R core.
He has
provided transformative energies
and software tools for data-analytic computing, emphasizing the
\mbox{opportunities} for interoperabilities among diverse components
\cite{nolan}.
Temple Lang's contribution to this collection
addresses the role of emerging
compilation techniques in the evolution of R. His paper
centers on the LLVM (formerly ``Low-Level Virtual Machine'')
compiler infrastructure as a basis for
transformation of R programs to highly performant
and retargetable\vadjust{\goodbreak} modules. This work is demonstrated
in two packages available through the Omegahat project.
The first is
\textit{Rllvm}, which connects R and LLVM infrastructures.
The second is
\textit{RllvmCompile}, which builds on top of \textit{Rllvm} to
transform selected R idioms to LLVM-based intermediate
representations, and ultimately to highly optimized
platform-targeted modules. The strategy is illustrated in a number
of practical examples.

Michael Lawrence and Martin Morgan are long-standing core
members of the Bioconductor project.
Their paper discusses software
design for analysis and visualization of genome-scale data, as
practiced at
Bioconductor \cite{gentGB}.
We are in a biological revolution with rapid advances
occurring in our understanding of how living organisms function. For
the statistics community R provides the golden standard for analysis:
open source, easy to contribute, hot-off-the-press methods,
high-level language, and good data plots. To analyze genomic data
in R requires data structures and algorithms that accommodate its
massive size and contextual structure. Additions to the Bioconductor suite
that better facilitate genomic data analysis are described conceptually and
illustrated with examples using currently available R packages.

Yihui Xie and Xiaoyue Cheng are major contributors to the Rstudio system,
providing highly accessible and extensible
methods for interactive graphics and
literate statistical computing \cite{knitr}.
Heike Hofmann is Professor of Statistics at Iowa State University
and is a widely recognized pioneer in visual inference with
high-dimensional data \cite{unwin,majum}.
The paper by Xie, Hofmann and Cheng
describes creating interactive graphics with mutable objects in R. Although
R has presented some especially important new ways to make
static plots of data, it was several steps back from
XLispStat \cite{Tierney} for interactive graphics.
Rudimentary R graphics are designed
from an ancient pen on paper model, which does not
easily enable the event loop hooks that are necessary to
make plots interactive and dynamic. Several new packages built on
new infrastructures in R enable mutable objects, objects that
can\vadjust{\goodbreak} changed inside of functions. How these mutable objects are
used to form the pipeline creating multiple linked interactive graphics
in the new \textit{cranvas} package is described by Xie, Hofmann and Cheng.

This collection of papers on statistical computing and visualization methods
is partly a product of invited session 267 of the
2012 Joint Statistical Meetings of the ASA, entitled ``Contemporary Software
\mbox{Design} Strategies for Statistical Methodologists.''
Subsequently, related symposia
on dynamic programming languages for analysis of large data have
been sponsored by NSF;
see \url{https://www.cs.purdue.edu/homes/jv/events/PBD13} and \url{http://www.ws13.dynali.org/talks.html}.

\section*{Acknowledgments}
Vincent Carey's contributions to this symposium were supported in part by National Institutes of Health Grant 5 U41 HG 004059 and NSF Grant DMS-1247813/IIS-124781.




\begin{thebibliography}{7}


\bibitem{gentGB}
\begin{bmisc}[auto:STB|2014/02/12|14:17:21]
\bauthor{\bsnm{Gentleman},~\bfnm{R.}\binits{R.}},
\bauthor{\bsnm{Carey},~\bfnm{V.}\binits{V.}},
\bauthor{\bsnm{Bates},~\bfnm{D.}\binits{D.}},
\bauthor{\bsnm{Bolstad},~\bfnm{B.}\binits{B.}},
\bauthor{\bsnm{Dettling},~\bfnm{M.}\binits{M.}},
\bauthor{\bsnm{Dudoit},~\bfnm{S.}\binits{S.}},
\bauthor{\bsnm{Ellis},~\bfnm{B.}\binits{B.}},
\bauthor{\bsnm{Gautier},~\bfnm{L.}\binits{L.}},
\bauthor{\bsnm{Ge},~\bfnm{Y.}\binits{Y.}},
\bauthor{\bsnm{Gentry},~\bfnm{J.}\binits{J.}},
\bauthor{\bsnm{Hornik},~\bfnm{K.}\binits{K.}},
\bauthor{\bsnm{Hothorn},~\bfnm{T.}\binits{T.}},
\bauthor{\bsnm{Huber},~\bfnm{W.}\binits{W.}},
\bauthor{\bsnm{Iacus},~\bfnm{S.}\binits{S.}},
\bauthor{\bsnm{Irizarry},~\bfnm{R.}\binits{R.}},
\bauthor{\bsnm{Leisch},~\bfnm{F.}\binits{F.}},
\bauthor{\bsnm{Li},~\bfnm{C.}\binits{C.}},
\bauthor{\bsnm{Maechler},~\bfnm{M.}\binits{M.}},
\bauthor{\bsnm{Rossini},~\bfnm{A.}\binits{A.}},
\bauthor{\bsnm{Sawitzki},~\bfnm{G.}\binits{G.}},
\bauthor{\bsnm{Smith},~\bfnm{C.}\binits{C.}},
\bauthor{\bsnm{Smyth},~\bfnm{G.}\binits{G.}},
\bauthor{\bsnm{Tierney},~\bfnm{L.}\binits{L.}},
\bauthor{\bsnm{Yang},~\bfnm{J.}\binits{J.}} \AND
\bauthor{\bsnm{Zhang},~\bfnm{J.}\binits{J.}}
(\byear{2004}).
\bhowpublished{Bioconductor: Open software development for computational biology and bioinformatics. \textit{Genome Biol.} \textbf{5} R80}.
\end{bmisc}
\bptok{imsref}%
\endbibitem

\bibitem{majum}
\begin{barticle}[mr]
\bauthor{\bsnm{Majumder},~\bfnm{Mahbubul}\binits{M.}},
\bauthor{\bsnm{Hofmann},~\bfnm{Heike}\binits{H.}} \AND
\bauthor{\bsnm{Cook},~\bfnm{Dianne}\binits{D.}}
(\byear{2013}).
\btitle{Validation of {v}isual {s}tatistical {i}nference,
{a}pplied to {l}inear {m}odels}.
\bjournal{J. Amer. Statist. Assoc.}
\bvolume{108}
\bpages{942--956}.
\bid{doi={10.1080/01621459.2013.808157}, issn={0162-1459}, mr={3174675}}
\end{barticle}
\bptok{imsref}%
\endbibitem

\bibitem{nolan}
\begin{bbook}[auto:STB|2014/02/12|14:17:21]
\bauthor{\bsnm{Nolan},~\bfnm{D.}\binits{D.}} \AND
\bauthor{\bsnm{Lang},~\bfnm{D.~T.}\binits{D.~T.}}
(\byear{2013}).
\btitle{XML and Web Technologies for Data Sciences with R}.
\bpublisher{Springer},
\blocation{New York}.
\end{bbook}
\bptok{imsref}%
\endbibitem

\bibitem{omega}
\begin{barticle}[mr]
\bauthor{\bsnm{Temple Lang},~\bfnm{Duncan}\binits{D.}}
(\byear{2000}).
\btitle{The {O}megahat environment: New possibilities for statistical computing}.
\bjournal{J. Comput. Graph. Statist.}
\bvolume{9}
\bpages{423--451}.
\bid{doi={10.2307/1390938}, issn={1061-8600}, mr={1818989}}
\end{barticle}
\bptok{imsref}%
\endbibitem

\bibitem{Tierney}
\begin{bbook}[auto:STB|2014/02/12|14:17:21]
\bauthor{\bsnm{Tierney},~\bfnm{L.}\binits{L.}}
(\byear{1990}).
\btitle{LISP-STAT: An Object-Oriented Environment for Statistical Computing and Dynamic Graphics}.
\bpublisher{Wiley},
\blocation{New York}.
\end{bbook}
\bptok{imsref}%
\endbibitem

\bibitem{unwin}
\begin{bbook}[auto:STB|2014/02/12|14:17:21]
\bauthor{\bsnm{Unwin},~\bfnm{A.}\binits{A.}},
\bauthor{\bsnm{Theus},~\bfnm{M.}\binits{M.}} \AND
\bauthor{\bsnm{Hofmann},~\bfnm{H.}\binits{H.}}
(\byear{2006}).
\btitle{Graphics of Large Datasets}.
\bpublisher{Springer},
\blocation{New York}.
\end{bbook}
\bptok{imsref}%
\endbibitem

\bibitem{knitr}
\begin{bbook}[auto:STB|2014/02/12|14:17:21]
\bauthor{\bsnm{Xie},~\bfnm{Y.}\binits{Y.}}
(\byear{2013}).
\btitle{Dynamic Documents with R and Knitr}.
\bpublisher{Chapman \& Hall/CRC},
\blocation{Boca Raton, FL}.
\end{bbook}
\bptok{imsref}%
\endbibitem

\end{thebibliography}
\end{document}